\documentclass[a4paper,12pt]{article}
\usepackage{amsmath,amsthm,amssymb,amsfonts}

\makeatletter

\@addtoreset{equation}{section}
\makeatother

\def\ep{\varepsilon}

\def\R{\mathbb R}
\def\S{\mathbb S}
\def\N{\mathbb N}
\def\pa{\partial}
\def\b{\backslash}

\def\diam{{\rm diam}(X)}

\begin{document}
\title{Stability for time-dependent inverse transport}

\author{Guillaume Bal and Alexandre Jollivet
        \thanks{Department of Applied Physics and 
        Applied Mathematics, Columbia University, 
        New York NY, 10027; gb2030@columbia.edu and aj2315@columbia.edu}}

\maketitle

\begin{abstract}
  This paper concerns the reconstruction of the absorption and
  scattering parameters in a time-dependent linear transport equation
  from full knowledge of the albedo operator at the boundary of a
  bounded domain of interest. We present optimal stability results on
  the reconstruction of the absorption and scattering parameters for a
  given error in the measured albedo operator.
\end{abstract}

\section{Introduction}

Inverse transport theory has many applications in e.g. medical and
geophysical imaging. It consists of reconstructing optical parameters
in a domain of interest from measurements of the transport solution at
the boundary of that domain. The optical parameters are the total
absorption (extinction) parameter $\sigma(x)$ and the scattering
parameter $k(x,v',v)$, which measures the probability of a particle at
position $x$ to scatter from direction $v'$ to direction $v$.

The domain of interest is probed as follows. A known flux of particles
enters the domain and the flux of outgoing particles is measured at
the domain's boundary. Several inverse theories may then be envisioned
based on available data. The least favorable situation is when the
density of outgoing particles is angularly averaged, which means that
only the spatial density of particles may be estimated and not the
phase space (space and direction) density.  Angular averaging may be
necessitated by equipment cost, time of acquisition of the
measurements, or low particle counts.  For uniqueness and stability
results in this setting, we refer the reader e.g. to Bal and Jollivet
[BJ2], Bal et al. [BLM], and Langmore [L].

A much more favorable situation is when the density of outgoing
particles is angularly resolved.  We may then be able to sample the
outgoing distribution of particles as a function of time if
sufficiently accurate equipment is available.  In many setting
however, only time independent measurements are feasible.

The uniqueness of the reconstruction of the optical parameters from
knowledge of angularly resolved measurements both in the
time-dependent and time-independent settings was proved in Choulli and
Stefanov [CS1, CS2]. We also refer the reader to Stefanov [S] for a
review of uniqueness results in inverse transport theory.  Stability
in the time-independent case has been analyzed in dimension $d=2,3$
under smallness assumptions for both optical parameters by Romanov
[R1, R2] and in dimension $d=2$ under smallness assumption for the
scattering parameter by Stefanov and Uhlmann [SU]. Partial results on
the stability of the reconstruction in the time-independent setting in
dimension $d=3$ were obtained in Wang [W] without smallness
assumptions.  Complete stability results in the time-independent case
in dimension $d\geq3$ were obtained by the authors in [BJ1]. The
present paper proves stability results for the time-dependent inverse
transport problem. We restrict ourselves to the case of elastic
scattering, where the velocity space may be modeled by the unit sphere
$\S^{d-1}$.  Optimal results on the stability of the optical
parameters are obtained in all dimensions $d\geq2$.

The rest of the paper is structured as follows. Section \ref{sec:fwd}
recalls useful results on the time-dependent linear transport
equation.  The main stability results of this paper are stated in
section \ref{sec:stab}. They are based on a decomposition of the
albedo operator used in [CS1] and recalled in section \ref{sec:dec}.
Useful regularity results on the decomposition are stated in
Proposition 3.2 and proved in section 4.  Our first stability result
is stated in Theorem 3.1. It shows how the Radon transform of the
absorption parameter and a weighted $L^1$ norm of the scattering
coefficient may be stably reconstructed from knowledge of the albedo
operator. Under additional regularity assumptions, Theorem 3.2 shows the
stability of the reconstruction of both optical parameters. Both
stability results are proved in section 5.

\section{The forward problem}
\label{sec:fwd}

In this section we introduce some notation and recall known facts
about the well-posedness of the forward transport problem.


\subsection{The linear Boltzmann transport equation}
Let $X$ be a bounded open subset of $\R^d$, $d\ge 2$, with a $C^1$ boundary $\pa X$. We denote the diameter of $X$ by $\diam$ ($\diam:=\sup_{(x,y)\in X^2}|x-y|$).
Let $\nu(x)$ denote the outward normal unit vector to $\pa X$ at $x\in \pa X$.
Let $\Gamma_{\pm}=\{(x,v)\in \pa X\times \S^{d-1}\ |\ \pm\nu(x)v>0\}$.
For $(x,v)\in \bar X\times \S^{d-1}$ we define $\tau_\pm (x,v)$ and $\tau(x,v)$ by 
$\tau_{\pm}(x,v):=\inf\{s\in (0,+\infty)\ |\ x\pm sv \not \in  X\}$ and $\tau(x,v):=\tau_-(x,v)+\tau_+(x,v)$.\\

Consider $\sigma:X\times \S^{d-1}\to \R$ and $k:X\times \S^{d-1}\times \S^{d-1}\to \R$ two nonnegative measurable functions.
We assume that $(\sigma,k)$ is admissible when
\begin{equation} \label{eq:hyp1p}
  \begin{array}{l}
0\le \sigma\in L^{\infty}(X\times \S^{d-1}),\\
0\le k(x,v',.)\in L^1(\S^{d-1})\textrm{ a.e. }(x,v')\in X\times \S^{d-1} \\
\sigma_p(x,v')=\displaystyle\int_{\S^{d-1}}k(x,v',v)dv\textrm{ belongs to } L^{\infty}(X\times \S^{d-1}).
\end{array}
\end{equation}

Let $T>\eta>0$. We consider the following linear Boltzmann transport equation with boundary conditions
\begin{equation}
{\pa u\over \pa t}(t,x,v)+v\nabla_xu(t,x,v)+\sigma(x,v)u(t,x,v)=\int_{\S^{d-1}}\!\!\!k(x,v',v)u(t,x,v')dv',\ (t,x,v)\in (0,T)\times X\times\S^{d-1},\label{B1}
\end{equation}
\begin{eqnarray}
&&u_{|(0,T)\times \Gamma_-}(t,x,v)=\phi(t,x,v),\nonumber\\
&&u(0,x,v)=0, \ (x,v)\in X\times \S^{d-1},\nonumber
\end{eqnarray}
where $\phi\in L^1((0,T),L^1(\Gamma_-,d\xi))$ and ${\rm supp} \phi\subseteq[0,\eta]$.

We assume here that scattering is elastic, which implies that the
speed of the particles is preserved by scattering while only the
direction of propagation may change. Elastic scattering is a good
approximation in many applications in medical and geophysical imaging.
Our results are stated for a (normalized) velocity space equal to the
unit sphere $\S^{d-1}$. Generalizations to other velocity spaces may
be obtained as in e.g. [BJ2] and [CS1, CS2].

\subsection{Semigroups and unbounded operators}
We introduce the following space
\begin{eqnarray}
{\mathcal Z}&:=&\{f\in L^1(X\times \S^{d-1})\ |\ v\nabla_xf\in L^1(X\times \S^{d-1})\},\,\,\label{B2a}\\
\|f\|_{\mathcal Z}&:=&\|f\|_{L^1(X\times \S^{d-1})}+\|v\nabla_x f\|_{L^1(X\times \S^{d-1})};\label{B2b}
\end{eqnarray}
where $v\nabla_x$ is understood in the distributional sense.

It is known [C1, C2] that the trace map $\beta_-$ from $C^1(\bar
X\times \S^{d-1})$ to $C(\Gamma_-)$ defined by
\begin{equation}
\beta_-(f)=f_{|\Gamma_-}
\end{equation}
extends to a continuous operator from ${\mathcal Z}$ onto $L^1(\Gamma_-, \tau_+(x,v)d\xi(x,v))$ and  admits a continuous lifting.
Note that $L^1(\Gamma_-,d\xi)$ is a subset of the space  $L^1(\Gamma_-, \tau_+(x,v)d\xi(x,v))$.\\

We introduce the following notation
\begin{eqnarray}
A_1f=-\sigma f,\ A_2f=\int_{\S^{d-1}}k(x,v',v)f(x,v')dv'.\label{B2c}
\end{eqnarray}
As $(\sigma,k)$ is admissible, the operators $A_1$ and $A_2$ are bounded operators in $L^1(X\times \S^{d-1})$.

Consider the following unbounded operators 
\begin{eqnarray} 
&&T_1f=-v\nabla_xf+A_1f,\ D(T_1)=\{f\in {\mathcal Z}\ |\ f_{|\Gamma_-}=0\},\label{B3}\\
&&Tf=T_1f+A_2 f,\ D(T)=D(T_1).
\end{eqnarray}
The unbounded operators $T_1$ and $T$ are generators of strongly
continuous semigroups $U_1(t)$ and $U(t)$, respectively, in
$L^1(X\times \S^{d-1})$ (see e.g. [DL, Proposition 2 p.226]).
In addition, $U_1(t)$ and $U(t)$ preserve the cone of positive
functions and $U_1(t)$ is given explicitly by the following formula
\begin{equation}
U_1(t)f=e^{-\int_0^t\sigma(x-sv,v)ds}f(x-tv,v)\theta(x-tv,x),\textrm{ for a.e. }(x,v)\in X\times \S^{d-1},\label{B4}
\end{equation}
for $f\in L^1(X\times \S^{d-1})$, where 
\begin{equation}
\theta(x,y)=
\left\lbrace
\begin{matrix}
1\textrm{ if }x+p(y-x)\in X \textrm{ for all } p\in [0,1],\\
0\textrm{ otherwise},
\end{matrix}
\right.\label{B4.0}
\end{equation}
for $(x,y)\in \R^d\times \R^d$.

We will use the Duhamel formula
\begin{equation}
U(r')= U_1(r')+\int_0^{r'}U_1(r'-s')A_2 U(s')ds',\textrm{ for }r'\ge 0.\label{C1}
\end{equation}

\subsection{Trace results}
We introduce the following space 
\begin{eqnarray}
{\mathcal W}&\!\!\!:=&\!\!\!\Big\{u\in L^1((0,T)\times X\times \S^{d-1})\ |\ \left({\pa \over \pa t}+v\nabla_x\right)u\in L^1((0,T)\times X\times \S^{d-1})\Big\},\,\,\,\,\,\,\,\,\,\,\,\,\,\,\,\,\,\label{B22a}\\
\|u\|_{\mathcal W}&\!\!\!:=&\!\!\!\|u\|_{L^1((0,T)\times X\times \S^{d-1})}+\left\|\left({\pa\over \pa  t}+v\nabla_x\right)u\right\|_{L^1((0,T)\times X\times \S^{d-1})};\label{B22b}
\end{eqnarray}
where ${\pa \over \pa t}$ and $v\nabla_x$ are understood in the distributional sense.

It is known [C1, C2] that the trace map $\gamma_-$ (respectively
$\gamma_+$) from $C^1([0,T]\times \bar X\times \S^{d-1})$ to
$C(X\times \S^{d-1})\times C((0,T)\times \Gamma_-)$ (respectively
$C(X\times \S^{d-1})\times C((0,T)\times \Gamma_+)$) defined by
\begin{equation}
\gamma_-(\psi)=(\psi(0,.), \psi_{|(0,T)\times \Gamma_-})\textrm{ (respectively }\gamma_+(\psi)=(\psi(T,.),\psi_{|(0,T)\times \Gamma_+})\textrm{)}
\end{equation}
extends to a continuous operator from ${\cal W}$ onto $L^1(X\times
\S^{d-1}, \tau_+(x,v)dx dv)\times L^1((0,T)\times \Gamma_-,
\min(T-t,\tau_+(x.v))dtd\xi(x,v))$ (respectively $L^1(X\times
\S^{d-1},\tau_-(x,v)dxdv)\times L^1((0,T)\times \Gamma_+,$

\noindent $\min(t,\tau_-(x,v))dtd\xi(x,v))$). In addition $\gamma_{\pm}$ admits a continuous lifting.
Note that $L^1(X\times \S^{d-1})$ is a subset of $L^1(X\times
\S^{d-1}, \tau_+(x,v)dx dv)$.  Note also that $L^1((0,T)\times
\Gamma_-,dt d\xi)$ (respectively $L^1((0,T)\times \Gamma_+,dt d\xi)$)
is a subset of $L^1((0,T)\times
\Gamma_-,\min(T-t,\tau_+(x.v))dtd\xi(x,v))$ (respectively
$L^1((0,T)\times \Gamma_+,\min(t,\tau_-(x,v))dtd\xi(x,v))$).\\

We now introduce the space 
\begin{equation}
W:=\{u\in {\mathcal W}\ |\ \gamma_-(u)\in L^1(X\times \S^{d-1})\times L^1((0,T)\times \Gamma_-,dt d\xi)\}.\label{B33}
\end{equation}
We recall the following trace results (see [C1, C2] in a more general
setting).  \vskip 2mm

{\bf Lemma 2.1.}
{\it The following equality is valid 
\begin{equation}
W=\{u\in {\mathcal W}\ |\ \gamma_+(u)\in L^1(X\times \S^{d-1})\times L^1((0,T)\times \Gamma_+,dt d\xi)\}.\label{B4a}
\end{equation}
In addition the trace maps 
\begin{eqnarray}
\gamma_{\pm}:W\to L^1(X\times \S^{d-1})\times L^1((0,T)\times \Gamma_\pm,dt d\xi)\nonumber\\
\textrm{ are continuous, onto, and admit continuous liftings.}\label{B4b}
\end{eqnarray}
}

\subsection{Solution to equation \eqref{B1}}
For any $r>0$, we identify the space $L^1((0,r),L^1(\Gamma_\pm,d\xi))$ with the space $L^1((0,r)\times\Gamma_\pm,dtd\xi)$, and we extend any function $\phi\in L^1((0,r), L^1(\Gamma_-,d\xi))$ by $0$ on $\R\b(0,r)$ (the extension is still denoted by $\phi$).\\

Let $\phi\in L^1((0,\eta),L^1(\Gamma_-,d\xi))$. We extend $\phi$ by $0$ outside $(0,\eta)$.
Then we consider the lifting  $G_-(t)\phi\in W$ of $(0,\phi)$  defined by
\begin{equation}
G_-(t)\phi(x,v):=e^{-\int_0^{\tau_-(x,v)}\sigma(x-sv,v)ds}\phi_-(t-\tau_-(x,v),x-\tau_-(x,v)v,v),\label{B5}
\end{equation}
for $(t,x,v)\in (0,T)\times X\times \S^{d-1}$.
Note that $G_-(.)\phi$ is a solution in the distributional sense of the equation $({\pa \over \pa t}+v\nabla_x)u+\sigma u=0$ in $(0,T)\times X\times \S^{d-1}$ and 
\begin{equation}
\|G_-(.)\phi\|_{\mathcal W}\le (1+\|\sigma\|_{\infty})\|G_-(.)\phi\|_{L^1((0,T)\times X\times \S^{d-1})}\le (1+\|\sigma\|_{\infty})T\|\phi_-\|_{L^1((0,\eta)\times\Gamma_-,dt d\xi)}.\label{B6}
\end{equation}
To prove the latter statements, one can use the change of variables
given by Lemma 4.1.  From \eqref{B6} we obtain that the map
$i:L^1((0,\eta), L^1(\Gamma_-,d\xi))\to W$ defined by
\begin{equation}
i(\phi)=G_-(.)\phi,\ \phi\in L^1((0,\eta), L^1(\Gamma_-,d\xi)),\label{B7}
\end{equation}
is continuous.\\

The following result holds (see [DL, Theorem 3 p. 229]).
\vskip 2mm
{\bf Lemma 2.2.} 
{\it The equation \eqref{B1} admits a unique solution $u$ in $W$ which is given by
\begin{equation}  
u(t)=G_-(t)\phi+\int_0^tU(t-s)A_2G_-(s)\phi ds.\label{B8}
\end{equation}
where $U(t)$ is the strongly continuous semigroup in $L^1(X\times \S^{d-1})$  introduced in section 2.2.
}\\

From \eqref{B7}, Lemma 2.2 and \eqref{B4b}, we obtain the existence of the albedo operator.
\vskip 2mm
{\bf Lemma 2.3.}
{\it The albedo operator  $A$ given by the formula
\begin{equation}
A\phi=u_{|(0,T)\times \Gamma_+}, \textrm{ for } \phi\in L^1((0,\eta),L^1(\Gamma_-,d\xi))\textrm{ where }u\textrm{ is given by }\eqref{B8}, 
\end{equation}
is well-defined and is a bounded operator from 
$L^1((0,\eta),L^1(\Gamma_-,d\xi))$ to $L^1((0,T),L^1(\Gamma_+,d\xi))$.
}

\section{Stability results for the inverse problem}
\label{sec:stab}

\subsection{Recall of uniqueness results}
\label{sec:rec}
Choulli-Stefanov [CS1] studied the uniqueness of the reconstruction of
$(\sigma,k)$ from the albedo operator by analyzing the distributional
kernel of that operator.  They considered the following problem
\begin{equation}
{\pa u\over \pa t}(t,x,v)+v\nabla_xu(t,x,v)+\sigma(x,v)u(t,x,v)=\int_{\S^{d-1}}k(x,v',v)u(t,x,v')dv',\ (t,x,v)\in \R\times X\times V,\label{S1}
\end{equation}
\begin{eqnarray}
&&u_{|\R\times \Gamma_-}(t,x,v)=\phi(t,x,v),\nonumber\\
&& u_{|t\ll0}=0,\nonumber
\end{eqnarray}
for $\phi\in L^1_{\rm comp}(\R,L^1(\Gamma_-,d\xi))$, where $V$ is an
open subset of $\R^d$, $d\ge 2$. The albedo operator is defined as an
operator from $L^1_{\rm comp}(\R,L^1(\Gamma_-,d\xi))$ to $L^1_{\rm
  loc}(\R,L^1(\Gamma_+,d\xi))$. They proved, in particular, that the
albedo operator uniquely determines the absorption and scattering
coefficient $(\sigma,k)$ provided that $\sigma$ is a function of $x$ and
$|v|$ only.  It is straightforward from the proof of this result (see
[CS1, Theorem 5.1, Propositions 5.1 and 5.2]) that the following
result holds.  \vskip 2mm

{\bf Proposition 3.1.}  {\it Assume that $(\sigma,k)$ are admissible
  and $\sigma(x,v)=f(x,|v|)$ for some real function $f$. Let
  $T>\eta>0$.  Then the following statements are valid:
\begin{itemize} 
\item[i] if $T>{\rm diam}(X)$  then the albedo operator \\$A:L^1((0,\eta),L^1(\Gamma_-,d\xi)) \to L^1((0,T),L^1(\Gamma_+,d\xi)) $ uniquely determines $\sigma$,
  
\item[ii] if $T>2{\rm diam}(X)$ then the albedo operator
  \\$A:L^1((0,\eta),L^1(\Gamma_-,d\xi)) \to
  L^1((0,T),L^1(\Gamma_+,d\xi)) $ uniquely determines $(\sigma,k)$,
\end{itemize}
}

In this paper we analyze the stability of the reconstruction of
$(\sigma,k)$ from the albedo operator. Our study is also based on the
distributional kernel of the albedo operator. In a first stage, we do
not assume that $\sigma(x,v)=f(x,|v|)$ for some real function $f$.

\subsection{Decomposition of the albedo operator}
\label{sec:dec}
Consider the distributional kernels 
\begin{equation}
\alpha_1(\tau,x,v,x',v')=e^{-\int_0^{\tau_-(x,v)}\sigma (x-sv,v)ds }\delta_v(v')\delta_{x-\tau_-(x,v)v}(x')\delta(\tau-\tau_-(x,v)),\label{D2a}
\end{equation}
\begin{eqnarray}
\alpha_2(\tau,x,v,x',v')&=&\int_0^{\tau_-(x,v)}e^{-\int_0^s\sigma(x-pv,v)ds-\int_0^{\tau_-(x-sv,v')}\sigma(x-t v-pv',v')dp}\label{D2b}\\
&&\times k(x-s v,v',v)\delta_{x-s v-\tau_-(x-s v,v')v'}(x')\delta(\tau-s-\tau_-(x-sv,v'))ds,\nonumber
\end{eqnarray}
for a.e. $(\tau,x,v,x',v')\in \R\times \Gamma_+\times \Gamma_-$ and
where we have defined $\int_{\S^{d-1}}f_1(v')\delta_v(v')dv'=f_1(v)$,
$\int_{\pa X}\delta_{x_0'}(x')f_2(x')d\mu(x')=f_2(x_0')$ and
$\int_{\R}\delta(\tau-s)f_3(\tau)d\tau=f_3(s)$ for $(v,x_0',s)\in
\S^{d-1}\times\pa X\times \R$ and for
$(f_1,f_2,f_3)\in C(\S^{d-1})\times C(\pa X)\times C(\R)$.\\
  
We consider the usual decomposition of the albedo operator as a sum of
three terms: the ballistic part (whose distributional kernel is given
by $\alpha_1$), the single scattering part (whose distributional
kernel is given by $\alpha_2$) and the multiple scattering (whose
distributional kernel is denoted by $\alpha_3$).  Using [CS1, Theorem
5.1], we know that $|\nu(x')v'|^{-1}\alpha_3\in L^\infty(\Gamma_-,
L_{\rm loc}^1(\R, L^1(\Gamma_+,d\xi)))$. The following Proposition 3.2
improves on the latter statement provided that $k\in L^\infty(X\times
\S^{d-1}\times \S^{d-1})$. The result will be used in the proof
of Theorem 3.1. \vskip 2mm

\noindent{\bf Proposition 3.2.} 
{\it Assume $d\ge 2$ and $(\sigma, k)$ admissible. Assume that $k\in L^\infty(X\times \S^{d-1}\times \S^{d-1})$. Then
\begin{equation}
A(\phi)(t,x,v):=\int_{(0,\eta)\times \Gamma_-}\hspace{-.75cm}
  (\alpha_1+\alpha_2+\alpha_3)
(t-t',x,v,x',v')\phi(t',x',v')dt'd\mu(x')dv',\,\,\,\,\label{D1}
\end{equation}
for a.e. $(t,x,v)\in (0,T)\times \Gamma_+$ and for any continuous and compactly supported function $\phi$ on $(0,\eta)\times \Gamma_-$, where
\begin{equation}
|\nu(x')v'|^{-1}\alpha_3\in L^\infty(\Gamma_-, L^p((-\eta,T), L^p(\Gamma_+,d\xi))), \quad \mbox{ for any } 1\le p< {d+1\over d}. \label{D3}
\end{equation}
}
Proposition 3.2 is proved in section 4.


\subsection{First stability result}

Now we assume that $X$ is a bounded open convex subset of $\R^d$,
$d\ge 2$, with $C^1$ boundary and that
\begin{equation}
  \label{eq:hyp2p}
  \begin{array}{l}
\textrm{the function }0\le\sigma \textrm{ is continuous and bounded on }X\times \S^{d-1},
\\
\textrm{the function }0\le k\textrm{ is continuous and bounded on }X\times \S^{d-1}\times \S^{d-1}.
  \end{array}
\end{equation}

Let $(\tilde \sigma, \tilde k)$ be a pair of absorption and scattering
coefficients that also satisfy \eqref{eq:hyp2p}.  Let $\tilde A$ be
the albedo operator from $L^1((0,\eta),L^1(\Gamma_-,d\xi))$ to
$L^1((0,T),L^1(\Gamma_+,d\xi))$ related to $(\tilde \sigma,\tilde k)$.

For $(x,v,s,w)\in \Gamma_\pm\times \R\times \S^{d-1}$,
$0<s<\tau_\mp(x,v)$, let $E_\mp(x,v,s,w)\ge 0$ be defined by
\begin{equation} 
E_\mp(x,v,s,w)=\exp\Big({-\int_0^s\sigma(x\mp pv,v)ds-\int_0^{\tau_\mp(x\mp sv,w)}\sigma(x\mp s v\mp p w,w)dp}\Big).\label{P-1}
\end{equation}
Replacing $\sigma$ by $\tilde \sigma$ in \eqref{P-1} we define $\tilde
E_\mp(x,v,s,w)$ similarly for $(x,v,s,w)\in \Gamma_\pm \times \R\times
\S^{d-1}$, $0<s<\tau_\mp(x,v)$.

Let $(x_0',v_0')\in \Gamma_-$.  For $\ep_1>0$ and $\ep_2>0$, let
$f_{\ep_1}\in C^1(\Gamma_-)$ and $g_{\ep_2}\in C^\infty(\R)$ be such
that
\begin{eqnarray}
&&f_{\ep_1}\ge 0,\ {\rm supp}f_{\ep_1}\subseteq \{(x',v')\in \Gamma_-\ |\ |x-x_0'|+|v'-v_0'|\le \ep_1\},\label{P0a}\\
&&\int_{\Gamma_-}f_{\ep_1}(x',v')d\xi(x',v')=1,\nonumber\\
&&g_{\ep_2}\ge 0,\ {\rm supp}g_{\ep_2}\subseteq (0,\min(\eta,\ep_2)),\ \int_0^{+\infty}g_{\ep_2}(t)dt=1.\label{P0b}
\end{eqnarray}
Consider the function $\phi_{\ep_1,\ep_2}\in C^1(\R\times \Gamma_-)$ defined by
\begin{equation}
\phi_{\ep_1,\ep_2}(t',x',v')=g_{\ep_2}(t')f_{\ep_1}(x',v'),\label{P0c}
\end{equation}
for $t'\in (0,+\infty)$ and $(x',v')\in \Gamma_-$. Note that ${\rm supp}\phi_{\ep_1,\ep_2}\subseteq (0,\eta)\times\Gamma_-$ (see \eqref{P0b}).
From \eqref{P0a} and \eqref{P0b} it follows that $|\nu(x')v'|\phi_{\ep_1,\ep_2}$ is a smooth approximation of the delta function on $\R\times \Gamma_-$ at $(0,x_0',v_0')$ as 
$\ep_1\to 0^+$ and $\ep_2\to 0^+$.

Let $\psi$ be any compactly supported continuous function on
$(0,T)\times\Gamma_+$ such that $\|\psi\|_{\infty}\le 1$.  First we
note that upon using the estimate $\|\psi\|_{\infty}\le 1$ and the
equality $\int_{(0,\eta)\times
  \Gamma_-}\phi_{\ep_1,\ep_2}(t,x,v)dtd\xi(x,v)=1$ we obtain that
\begin{eqnarray}
\left|\int_{(0,T)\times \Gamma_+}\psi(t,x,v) \left(A-\tilde A\right)\phi_{\ep_1,\ep_2}(t,x,v)dt d\xi(x,v)\right|&\le& 
\|(A-\tilde A)\phi_{\ep_1,\ep_1}\|_{L^1((0,T),L^1(\Gamma_+,d\xi))}\nonumber\\
&\le& \|A-\tilde A\|_{\eta,T},\label{P0d}
\end{eqnarray}
where $\|.\|_{\eta,T}:=\|.\|_{{\cal
    L}(L^1((0,\eta),L^1(\Gamma_-,d\xi)),L^1((0,T),L^1(\Gamma_+,d\xi)))}$.

In addition, it follows from \eqref{D1}--\eqref{D3} that for any
compactly supported continuous function $\psi$ on
$(0,T)\times\Gamma_+$, we have
\begin{equation}
\int_{(0,T)\times \Gamma_+}\psi(t,x,v) \left(A-\tilde A\right)\phi_{\ep_1,\ep_2}(t,x,v)dt d\xi(x,v)=I_1(\psi,\ep_1,\ep_2)+I_2(\psi,\ep_1,\ep_2)+I_3(\psi,\ep_1,\ep_2),\label{P0e}
\end{equation}
where
\begin{eqnarray}
I_1(\psi,\ep_1,\ep_2)&=&\int_0^T\int_{\Gamma_+}\left(e^{-\int_0^{\tau_-(x,v)}\sigma (x-sv,v)ds }-e^{-\int_0^{\tau_-(x,v)}\tilde\sigma (x-sv,v)ds }\right)\nonumber\\
&&\times\psi(t,x,v)\phi_{\ep_1,\ep_2}(t-\tau_-(x,v),x-\tau_-(x,v)v,v)d\xi(x,v)dt\label{P0f}\\
I_2(\psi,\ep_1,\ep_2)&=&\int_0^T\int_{\Gamma_+}\psi(t,x,v)\int_{\S^{d-1}}\int_0^{\tau_-(x,v)}\Big(E_-(x,v,s,v')k(x-s v,v',v)\nonumber\\
\nonumber\\
&&\left.-\tilde E_-(x,v,s,v')\tilde k(x-sv,v',v)\right) \psi_{\ep_1,\ep_2}(t-s-\tau_-(x-sv,v'),\nonumber\\
&&x-s v-\tau_-(x-s v,v')v',v')ds dv'd\xi(x,v)dt\label{P0g}\\
|I_3(\psi,\ep_1,\ep_2)|&\le&C\left(\int_0^T\int_{\Gamma_-}|\psi(t,x,v)|^{p'}d\xi(x,v)dt\right)^{1\over p'},\label{P0gg}
\end{eqnarray}
where $C$ is a constant that does not depend on $\ep_1$, $\ep_2$ (for \eqref{P0gg} we also used H\"older inequality and the equality 
$\|\phi_{\ep_1,\ep_2}\|_{L^1((0,\eta), L^1(\Gamma_-,d\xi))}=1$).\\

Using \eqref{P0a}--\eqref{P0c}, \eqref{eq:hyp2p}, we obtain the following preparatory Lemma 3.1.
\vskip 2mm
\noindent{\bf Lemma 3.1.}
{\it Assume that $X$ is convex and that $(\sigma,k)$ and $(\tilde \sigma,\tilde k)$ both satisfy \eqref{eq:hyp2p}. Then the following statements are valid:
\begin{itemize}

\item[i.] if $T>\diam$ then
\begin{eqnarray}
\lim_{\ep_2\to 0^+}\lim_{\ep_1\to 0^+}I_1(\psi,\ep_1,\ep_2)&\!\!\!=\!\!\!&\psi(\tau_+(x_0',v_0'),x_0'+\tau_+(x_0',v_0')v_0',v_0')\label{P0h}\\
&&\!\!\times\!\left(e^{-\int_0^{\tau_+(x_0',v_0')}\sigma(x_0'+sv_0',v_0')ds}-e^{-\int_0^{\tau_+(x_0',v_0')}\tilde \sigma(x_0'+sv_0',v_0')ds}\!\right)\!,
 \!\!\!  \nonumber
\end{eqnarray}
for any compactly supported and continuous function $\psi$ on $(0,T)\times\Gamma_+$;

\item[ii.] if $T>2\diam$ then
\begin{equation}
\lim_{\ep_2\to 0^+}\lim_{\ep_1\to 0^+}I_2(\psi,\ep_1,\ep_2)=I_2^1(\psi)+I_2^2(\psi),\label{P0i}
\end{equation}
for any compactly supported and continuous function $\psi$ on $(0,T)\times\Gamma_+$, where
\begin{eqnarray}
I_2^1(\psi)&=&\int_{\S^{d-1}}\int_0^{\tau_+(x_0',v_0')}\psi(s+\tau_+(x_0'+sv_0',v),x_0'+sv_0'+\tau_+(x_0'+sv_0',v)v,v)\nonumber\\
&&(k-\tilde k)(x_0'+sv_0',v_0',v)E_+(x_0',v_0',s,v)dsdv,\label{P0ja}\\
I_2^2(\psi)&=&\int_{\S^{d-1}}\int_0^{\tau_+(x_0',v_0')}\psi(s+\tau_+(x_0'+sv_0',v),x_0'+sv_0'+\tau_+(x_0'+sv_0',v)v,v)
\nonumber\\
&&\tilde k(x_0'+sv_0',v_0',v)(E_+-\tilde E_+)(x_0',v_0',s,v)dsdv,\label{P0jb}
\end{eqnarray}
where $E_+$ and $\tilde E_+$ are defined by \eqref{P-1}.
\end{itemize}
}

\vskip 2mm
Lemma 3.1 is proved in section 5.\\

Taking account of Lemma 3.1 and \eqref{P0d}, and choosing an appropriate sequence of functions ``$\psi$'', we obtain the main result of this paper:
\vskip 2mm
\noindent{\bf Theorem 3.1.}
{\it Let $T>\eta>0$. Assume that $d\ge 2$ and $X$ is convex and $(\sigma,k)$ and $(\tilde \sigma, \tilde k)$ both satisfy condition \eqref{eq:hyp2p}. Then the following 
statements are valid:
\begin{itemize}

\item[i.] if $T>\diam$, then
\begin{equation}
\label{P0k}\left|\exp\Big({-\int_0^{\tau_+(x_0',v_0')}\hspace{-.5cm}\sigma(x_0'+sv_0',v_0')ds}\Big)-\exp\Big({-\int_0^{\tau_+(x_0',v_0')}\hspace{-.5cm}\tilde\sigma(x_0'+sv_0',v_0')ds}\Big)\right|
\le \|A-\tilde A\|_{\eta,T}; 
\end{equation}

\item[ii.] if $T>2\diam$, then
\begin{eqnarray}
&&\int_{\S^{d-1}}\int_0^{\tau_+(x_0',v_0')}\left|k-\tilde k\right|(x_0'+sv_0',v_0',v)E_+(x_0',v_0',s,v)dsdv\nonumber\\
&&\le \tau_+(x_0',v_0')\sup_{s\in (0,\tau_+(x_0',v_0'))}\tilde \sigma_p(x_0'+sv_0',v_0') \sup_{s\in (0,\tau_+(x_0',v_0'))\atop v\in \S^{d-1}}\left|E_+-\tilde E_+\right|(x_0',v_0',s,v)\nonumber\\
&&+\|A-\tilde A\|_{\eta,T},\label{P0l}
\end{eqnarray}
\end{itemize}
where  $\|.\|_{\eta,T}:=\|.\|_{{\cal L}(L^1((0,\eta),L^1(\Gamma_-,d\xi)),L^1((0,T),L^1(\Gamma_+,d\xi)))}$ and where $E_+$ and $\tilde E_+$ are defined by \eqref{P-1}. 
}
\vskip 2mm
The proof of Theorem 3.1 is given in section 5.\\

{\bf Remark 3.1.} One can prove that estimate \eqref{P0k} still holds
a.e. $(x_0',v_0')\in \Gamma_-$ provided that $T>\diam$ and $k\in
L^\infty(X\times \S^{d-1}\times \S^{d-1})$ and without assuming
\eqref{eq:hyp2p} nor that $X$ is convex.

\subsection{Second stability result} 
We now impose that the absorption coefficient $\sigma$ does not depend
on the velocity variable, i.e. $\sigma(x,v)=\sigma(x),$ $x\in X$.
Then let
\begin{eqnarray}
{\mathcal M}&:=&\big\{(\sigma(x), k(x,v',v)) \in L^\infty(X)\times L^\infty(X\times \S^{d-1}\times \S^{d-1}) \ |\ (\sigma,k)\textrm{ satisfies } \eqref{eq:hyp2p}, 
\nonumber\\
&&\textrm{and } \sigma\in H^{{d\over 2}+\tilde r}(X), \|\sigma\|_{H^{{d\over 2}+\tilde r}(X)}\le M, \|\sigma_p\|_{\infty}\le M
\big\},\label{5.1}
\end{eqnarray}
for some $\tilde r>0$ and $M>0$.  Using Theorem 3.1 for
any $(x_0',v_0')\in \Gamma_-$ we obtain the following Theorem 3.2.
\vskip 2mm
\noindent{\bf Theorem 3.2.}
{\it Assume that $d\ge 2$ and $X$ is convex. Let $T>\eta>0$. For any
  $(\sigma,k)\in {\mathcal M}$ and $(\tilde \sigma, \tilde k)\in
  {\mathcal M}$ the following stability estimates are valid:
\begin{itemize}

\item[i.]if $T>\diam$ then
\begin{equation}
\|\sigma-\tilde\sigma\|_{H^s(X)}\le C_1\|A-\tilde A\|_{\eta,T}^\kappa,\label{t5.2a}
\end{equation}
where ${-{1\over 2}}\le s<{d\over 2}+\tilde r$, $\kappa={d+2(\tilde r-s)\over d+1+2\tilde r}$, and $C_1=C_1(X, M, s, \tilde r)$; 

\item[ii.]if $T>2\diam$ then
\begin{eqnarray}
&&\int_{\S^{d-1}}\int_0^{\tau_+(x_0',v_0')}\left|k(x_0'+s'v_0',v_0',v)-\tilde k(x_0'+s'v_0',v_0',v)\right|ds'dv\nonumber\\&\le&
C_2\|A-\tilde A\|_{\eta,T}^\kappa\left(1+\|A-\tilde A\|_{\eta,T}^{1-\kappa}\right),\label{t5.2b}
\end{eqnarray}
for $(x_0', v_0')\in \Gamma_-$, and
where $\kappa={2(\tilde r-r)\over d+1+2 \tilde r}$, $0< r<\tilde r$, and $C_2=C_2(X, M, r, \tilde r)$;

\item[iii.]in addition, if $T>2\diam$ then
\begin{equation}
\|k-\tilde k\|_{L^1(X\times \S^{d-1}\times \S^{d-1})}\le C_3\|A-\tilde A\|_{\eta,T}^\kappa\left(1+\|A-\tilde A\|_{\eta,T}^{1-\kappa}\right),\label{t5.2c}
\end{equation}
where $\kappa={2(\tilde r-r)\over d+1+2 \tilde r}$, $0< r<\tilde r$, and $C_3=C_3(X, M, r, \tilde r)$.

\end{itemize}
}

\vskip 1mm Theorem 3.2 is proved in section 5.  \vskip 1mm {\bf Remark
  3.2.}  Stability estimates similar to \eqref{t5.2a} were given by
Cipolatti-Motta-Roberty [CMR, Theorem 1.1].  They proved \eqref{t5.2a}
for $s=-{1\over 2}$ under the assumptions $k$, $\tilde k\in
L^\infty(X,L^2(\S^{d-1}\times \S^{d-1}))$,
$\max(\|\sigma\|_{\infty},\|\tilde \sigma\|_\infty) \le M$ (and
$\max(\|\sigma_p\|_\infty,\|\tilde \sigma_p\|_\infty)<\infty$). They
also proved \eqref{t5.2a} for $0<s<\tilde r$ under the assumptions
$k$, $\tilde k\in L^\infty(X,L^2(\S^{d-1}\times \S^{d-1}))$,
$\sigma,\tilde \sigma\in H^{{d\over 2}+\tilde r}(X)$ and
$\max(\|\sigma\|_{H^{{d\over 2}+\tilde r}},
\|\tilde\sigma\|_{H^{{d\over 2}+\tilde r}})\le M$.

\section{Proof of Proposition 3.2}

Before giving the proof of Proposition 3.2, we need Lemmas 4.1,
4.2, 4.3.  \vskip 2mm

\noindent{\bf Lemma 4.1.}
{\it For $f\in L^1(X\times \S^{d-1})$ we have 
\begin{equation}
\int_{X\times \S^{d-1}}f(x,v)dx dv=\int_{\Gamma_\pm}\int_0^{\tau_{\mp}(x,v)}f(x\mp sv,v)dsd\xi(x,v).\label{D4}
\end{equation}
}
\vskip 2mm
For the proof of Lemma 4.1, see [CS2, Lemma 2.1].\\

Let $m\ge 1$. For $U$ a subset of $\R^m$, we denote by $\chi_U$ the
function from $\R^m$ to $\R$ defined by
\begin{equation}
\chi_U(x)=
\left\lbrace
\begin{matrix}
1\textrm{ if } x\in U,\\
0\textrm{ otherwise.}\end{matrix}
\right.
\label{D4-5}
\end{equation}
\\
{\bf Lemma 4.2.}  {\it Let $T>0$ and let $1<p<{d+1\over d}$.  Consider
  the nonnegative measurable function $\beta:X\to \R$ defined by
\begin{equation}
\beta(x')=\int_{\Gamma_+}\int_{|x-x'|}^{T+\diam}{(s-(x-x')v)^{p(d-3)}\over |x-x'-sv|^{p(2d-4)}}ds d\xi(x,v)\label{D5}
\end{equation}
for a.e. $x'\in X$.
Then 
\begin{equation}
\beta\in L^\infty(X).\label{D6}
\end{equation}
}
\begin{proof}[Proof of Lemma 4.2.] We first consider the case $d=2$.

We have
\begin{eqnarray*}
\beta(x')&\le&\int_{\pa X}\int_{|x-x'|}^{+\infty}\int_{\S^1}
{1\over (s-(x-x')v)^p} dv ds d\mu(x)\\
&\le&\int_{\pa X}\int_0^{2\pi}\int_{|x-x'|}^{+\infty}{-1\over p-1}{d\over ds}{1\over (s-|x-x'|\cos\omega)^{p-1}} d\omega ds d\mu(x)\\
&=&{1\over p-1}\int_{\pa X}{1\over |x-x'|^{p-1}}dx \int_0^{2\pi}{1\over (1-\cos\omega)^{p-1}}d\omega,
\end{eqnarray*}
for a.e. $x'\in X$. Hence using the estimate $p<1+{1\over 2}$, we obtain
\begin{equation}
\|\beta\|_{L^\infty( X)}\le {1\over p-1}\int_0^{2\pi}{1\over (1-\cos\omega)^{p-1}}d\omega\sup_{z\in X}\int_{\pa X}{1\over |x-z|^{p-1}}d\mu (x)<\infty.\label{C25b}
\end{equation}

Now assume $d=3$.  Using \eqref{D5}, spherical coordinates and
performing the change of variables ``$s$'' $=|x-x'|s$, we obtain
\begin{eqnarray*}
\beta(x')&\le&\int_{\pa X}\int_{\S^2}\int_{|x-x'|}^{+\infty}{1\over |sv-(x-x')|^{2p}}ds dv d\mu(x)\\
&=&2\pi\int_{\pa X}{1\over |x-x'|^{2p-1}}d\mu(x)\int_1^\infty{1\over 2s(p-1)}\int_{-{\pi \over 2}}^{\pi\over 2}{d\over d\omega}{1\over(s^2+1-2s\sin\omega)^{p-1}}d\omega ds\\
&\le&2\pi\int_{\pa X}{1\over |x-x'|^{2p-1}}d\mu(x)\int_1^\infty{1\over 2(p-1)s(s-1)^{2p-2}}ds.
\end{eqnarray*}
Therefore using the estimate $1<p<1+{1\over 3}$, we obtain
\begin{equation}
\|\beta\|_{L^\infty(X)}\le 2\pi\sup_{z\in X}\int_{\pa X}{1\over |x-z|^{2p-1}}d\mu(x)\int_1^\infty{1\over 2(p-1)s(s-1)^{2p-2}}ds<\infty.\label{C25a}
\end{equation}

Finally assume $d\ge 4$.  Note that $|s-v(x-x')|=|(sv-(x-x'))v|\le
|sv-(x-x')|$ for $s\in \R$ and $x,$ $x'\in \R^d$. Using in particular
the latter estimate and \eqref{D5}, we obtain
\begin{eqnarray*}
\beta(x')&\le&\int_{\pa X}\int_{\S^{d-1}}\int_{|x-x'|}^{+\infty}{(s-v(x-x'))\over |sv-(x-x')|^{p(d-1)+1}}ds dv d\mu(x)\\
&=&{1\over (p(d-1)-1)}\int_{\pa X}\int_{\S^{d-1}}{1\over (||x-x'|v-(x-x')|)^{p(d-1)-1}} dv d\mu(x)\\
&=&{\rm Vol}(\S^{d-2})\int_{\pa X}{1\over (p(d-1)-1)(|x-x'|\sqrt{2})^{p(d-1)-1}}d\mu(x)\int_{-{\pi \over 2}}^{\pi\over 2}{\cos\omega^{d-2}\over (1-\sin \omega)^{p(d-1)-1\over 2}}d\omega,
\end{eqnarray*}
for a.e. $x'\in X$. Hence using the estimate $p<1+{1\over d}$, we obtain
\begin{eqnarray}
\|\beta\|_{L^\infty(X)}&\le&{\rm Vol}(\S^{d-2}) {1\over (p(d-1)-1)2^{p(d-1)-1\over 2}}\sup_{z\in X}\int_{\pa X}{1\over |x-z|^{p(d-1)-1}}d\mu(x)\nonumber\\
&&\times \int_{-{\pi \over 2}}^{\pi\over 2}{\cos\omega^{d-2}\over 
(1-\sin \omega)^{p(d-1)-1\over 2}}d\omega\,\,<\,\,\infty.\label{C25c}
\end{eqnarray}

\end{proof}

Finally, we need the following Lemma 4.3.  \vskip 2mm
\noindent{\bf Lemma 4.3.}
{\it Consider the nonnegative measurable function $\gamma:(0,T)\times X\times \S^{d-1}\times X\times \S^{d-1}\to \R$ defined by
\begin{eqnarray}
&&\gamma(t,x,v,x',v')=2^{d-2}\chi_{(0,t)}(|x-x'|)\left[e^{-\int_0^{s_1}\sigma(x-sv,v)ds-\int_0^{t-s_1}\sigma(x-s_1v-pv_1,v_1)dp}\theta(x,x-s_1v)\right.\nonumber\\
&&\times\left.\theta(x-s_1v,x')k(x-s_1v,v_1,v)k(x',v',v_1)\right]_{s_1={t^2-(x-x')^2\over 2(t-(x-x')v)}}{(t-(x-x')v)^{d-3}\over |tv-x-x'|^{2d-4}},\label{D7}
\end{eqnarray}
where $\theta$ is defined by \eqref{B4.0}. Then
\begin{equation}
\left(\int_0^tU_1(t-s_1)A_2U_1(s_1)A_2f ds_1\right) (x,v)=\int_{X\times \S^{d-1}}\gamma(t,x,v,x',v')f(x',v')dx'dv'\label{D8}
\end{equation}
for $t\in (0,T)$ and for a.e. $(x,v) \in X\times \S^{d-1}$ and  for $f\in L^1(X\times \S^{d-1})$.
}

\begin{proof}[Proof of Lemma 4.3]
Let $t\in (0,T)$ and let $f\in L^1(X\times \S^{d-1})$.
From \eqref{B4} and \eqref{B2c},  it follows that
\begin{eqnarray}
\int_0^tU_1(t-s_1)A_2U_1(s_1)A_2f ds_1&\!\!\!=\!\!\!&\int_0^t\int_{\S^{d-1}\times \S^{d-1}}\!\!\!\!\!\!\!\!\!e^{-\int_0^{t-s_1}\sigma(x-pv,v)dp-\int_0^{s_1}\sigma(x-(t-s_1)v-pv_1,v_1)dp}\nonumber\\
&&\times k(x-(t-s_1)v,v_1,v)k(x-(t-s_1)v-s_1v_1,v',v_1)\nonumber\\
&&\times\theta(x-(t-s_1)v,x)\theta(x-(t-s_1)v-s_1v_1,x-(t-s_1)v)\nonumber\\
&&\times f(x-(t-s_1)v-s_1v_1,v')dv'dv_1ds_1.\label{C7}
\end{eqnarray}
Performing the change of variables ``$s_1=t-s_1$'' and then performing
the change of variables ``$x'=x-(t-s_1)v_1-s_1v$''
($2^{d-2}{\left((tv-(x-x'))v\right)^{d-3} \over
  |x-x'-tv|^{2d-4}}dx'=dv_1ds_1$), we obtain \eqref{D8}.
\end{proof}

\begin{proof}[Proof of Proposition 3.2]
  Let $\phi \in L^1((0,\eta), L^1(\Gamma_-,d\xi))$. Let $u$ be the
  solution of \eqref{B1}. Using twice Duhamel's formula \eqref{C1} and
  using \eqref{B8} we obtain
\begin{equation}
u(t)=R_1(t)+R_2(t)+R_3^1(t)+R_3^2(t),\label{C2}
\end{equation}
for $t\in (0,T)$ where
\begin{eqnarray}
R_1(t)&=&G_-(t)\phi,\label{C3a}\\
R_2(t)&=&\int_0^t U_1(t-t')A_2G_-(t')\phi dt', \label{C3b}\\
R_3^1(t)&=&\int_0^t\int_0^{t-t'}U_1(t-t'-s_1)A_2 U_1(s_1)A_2G_-(t')\phi ds_1dt',\label{C3c}\\
R_3^2(t)&=&\int_0^t\int_0^{t-t'}\int_0^{t-t'-s_2} U_1(t-t'-s_2-s_1)A_2 U_1(s_1)A_2 \label{C3d}\\
&&U(s_2)A_2G_-(t')\phi ds_1ds_2dt'.\nonumber
\end{eqnarray}

From \eqref{C3a} and \eqref{B5}, it follows that
\begin{eqnarray}
{R_1}_{|(0,T)\times \Gamma_+}(t,x,v)&=&e^{-\int_0^{\tau_-(x,v)}\sigma(x- sv,v)ds}\phi(t-\tau_- (x,v), x- \tau_-(x,v)v,v)\nonumber\\
&=&\int_{(0,\eta)\times \Gamma_-}\hspace{-.5cm}\alpha_1(t-t',x,v,x',v')\phi(t',x',v')dt' d\xi(x',v'),\label{C4}
\end{eqnarray}
where $\alpha_1$ is defined by \eqref{D2a}.

From \eqref{B4}, \eqref{B5} and \eqref{C3b}, it follows that
\begin{eqnarray} 
R_2(t,x,v)&=&\int_0^t\theta(x-t'v,x)\int_{\S^{d-1}}k(x-t'v,v',v)e^{-\int_0^{t'}\sigma(x-pv,v)dp-\int_0^{\tau_-(x-t'v,v')}\sigma(x-t'v-pv',v')dp}\nonumber\\
&&\times \phi(t-t'-\tau_-(x-t'v,v'),x-t'v-\tau_-(x-t'v,v')v',v')dv' dt'.\nonumber
\end{eqnarray}
Hence
\begin{equation}
{R_2}_{|(0,T)\times \Gamma_-}(t,x,v)=\int_{(0,\eta)\times \Gamma_-}\alpha_2(t-t',x,v,x',v')\phi(t',x',v')dt' d\xi(x',v'),\label{C5}
\end{equation}
where $\alpha_2$ is defined by \eqref{D2b}.

From \eqref{C3c} and \eqref{D8} (with ``$t$''$=t-t'$), it follows that
\begin{equation}
R_3^1(t)=\int_{(0,+\infty)\times X\times \S^{d-1}}\chi_{(0,+\infty)}(t-t')\gamma(t-t',x,v,x',v')G_-(t')\phi(x',v')dt'dx'dv',\label{C6a}
\end{equation}
where $\gamma$ is defined by \eqref{D7}.
Using \eqref{C6a} and \eqref{D7} we obtain
\begin{equation}
{R_3^1}_{|(0,T)\times \Gamma_+}(t,x,v)=\int_{(0,T)\times X\times \S^{d-1}}\tilde{\alpha_3}^1(t-t',x,v,x',v')(G_-(t')\phi)(x',v')dt'dx'dv',\label{C9}
\end{equation}
where
\begin{eqnarray}
&&\tilde{\alpha_3}^1(\tau,x,v,x',v')=\chi_{(0,+\infty)}(\tau-|x-x'|)\left[e^{-\int_0^{s_1}\sigma(x-pv,v)dp-\int_0^{\tau-s_1}\sigma(x-s_1v-pv_1,v_1)dp}\right.\nonumber\\
&&\times \left.k(x-s_1v,v_1,v)k(x',v',v_1)\theta(x,x-s_1v)\theta(x-s_1v,x')\right]_{s_1={\tau^2-|x-x'|^2\over 2(\tau-(x-x')v)}\atop v_1={x-x'-s_1v\over \tau-s_1}}\nonumber\\
&&\times 2^{d-2}{\left(\tau-(x-x')v\right)^{d-3}\over |x-x'-\tau v|^{2d-4}}\label{C10}
\end{eqnarray}
for a.e. $(\tau,x,v,x',v')\in \R\times \Gamma_+\times X\times\S^{d-1}$.
 
From \eqref{C3d} and \eqref{D8} it follows that
\begin{equation}
R_3^2(t)=\int_0^t\int_0^{t-t'}\int_{X\times\S^{d-1}}\gamma(t-t'-s_2,x,v,x',v')(U(s_2)A_2G_-(t')\phi)(x',v')dx'dv'ds_2dt',\label{C11}
\end{equation}
where $\gamma$ is defined by \eqref{D7}.
Hence
\begin{equation}
{R_3^2}_{|(0,T)\times \Gamma_+}(t,x,v)=\int_0^t\int_0^{t-t'}\int_{X\times\S^{d-1}}\tilde \gamma(t-t'-s_2,x,v,x',v')(U(s_2)A_2G_-(t')\phi)(x',v')dx'dv'ds_2dt',\label{C11b}
\end{equation}
for a.e. $(t,x,v)\in (0,T)\times\Gamma_+$ where
\begin{eqnarray}
&&\tilde\gamma(r,x,v,x',v')=2^{d-2}\chi_{(0,r)}(|x-x'|)\left[e^{-\int_0^{s_1}\sigma(x-sv,v)ds-\int_0^{r-s_1}\sigma(x-s_1v-pv_1,v_1)dp}\theta(x,x-s_1v)\right.\nonumber\\
&&\times\left.\theta(x-s_1v,x')k(x-s_1v,v_1,v)k(x',v',v_1)\right]_{s_1={r^2-(x-x')^2\over 2(r-(x-x')v)}}{(r-(x-x')v)^{d-3}\over |rv-x-x'|^{2d-4}},\label{C12}
\end{eqnarray}
for a.e. $(r,x,v,x',v')\in (0,T)\times \Gamma_+\times X\times\S^{d-1}$.

Let $\psi\in L^\infty((0,T)\times\Gamma_+)$. Assume that $k\in L^\infty(X\times \S^{d-1}\times \S^{d-1})$.
From \eqref{C9} it follows that
\begin{eqnarray}
&&\left|\int_0^T\int_{\Gamma_+}\psi(t,x,v){R_3^1}_{|(0,T)\times \Gamma_+}(t,x,v)dtd\xi(x,v)\right|\nonumber\\
&&=\left|\int_{(0,T)\times X\times \S^{d-1}}(G_-(t')\phi)(x',v')\int_0^T\int_{\Gamma_+}\tilde{\alpha_3}^1(t-t',x,v,x',v')\psi(t,x,v)d\xi(x,v)dt dt' dx'dv'\right|\nonumber
\end{eqnarray}
\begin{equation}
\le \|G_-(.)\phi\|_{L^1((0,T)\times X\times \S^{d-1})}\left\|\int_0^T\int_{\Gamma_+}\tilde{\alpha_3}^1(t-t',x,v,x',v')\psi(t,x,v)d\xi(x,v)dt dt'
\right\|_{L^\infty(\R_{t'}\times X_{x'}\times \S^{d-1}_{v'})}.
\label{C15}
\end{equation}

From Lemma 4.1 and \eqref{B5}, it follows that
\begin{equation}
\|G_-(t')\phi\|_{L^1(X\times \S^{d-1})}\le \|\phi\|_{L^1((0,\eta),L^1(\Gamma_-,d\xi))},\textrm{ for }t'\in (0,T).\label{C15a}
\end{equation}
From \eqref{C10}, H\"older's inequality and Lemma 4.2, it follows that
\begin{eqnarray}
&&\left|\int_0^T\int_{\Gamma_+}\tilde{\alpha_3}^1(t-t',x,v,x',v')\psi(t,x,v)d\xi(x,v)dt\right|\nonumber\\
&&\le\|k\|_{\infty}^2\int_0^T\int_{\Gamma_+}
 {\chi_{(0,+\infty)}(t-t'-|x-x'|)(t-t'-(x-x')v)^{d-3}\over 2^{2-d}|x-x'-(t-t')v|^{2d-4}}|\psi|(t,x,v)d\xi(x,v) dt\nonumber\\
&&\le 2^{d-2}\|k\|_{\infty}^2\left(\int_0^T\int_{\Gamma_+}|\psi(t,x,v)|^{p'}d\xi(x,v)dt\right)^{1\over p'}\beta(x')^{1\over p}\nonumber\\
&&\le 2^{d-2}\|k\|_{\infty}^2\|\beta\|_{L^\infty(X)}^{1\over p}\left(\int_0^T\int_{\Gamma_+}|\psi(t,x,v)|^{p'}d\xi(x,v)dt\right)^{1\over p'}, \label{C16}
\end{eqnarray}
for a.e. $(t',x',v')\in (0,T)\times X\times \S^{d-1}$.

Using \eqref{C15}--\eqref{C16} we obtain
\begin{eqnarray}
&&\left|\int_0^T\int_{\Gamma_+}\psi(t,x,v){R_3^1}_{|(0,T)\times \Gamma_+}(t,x,v)dtd\xi(x,v)\right|\label{C16b}\\
&&\le 2^{d-2}T\|k\|_{\infty}^2\|\beta\|_{L^\infty(X)}^{1\over p}\|\phi\|_{L^1((0,\eta),L^1( \Gamma_-,d\xi))}\left(\int_0^T\int_{\Gamma_+}|\psi(t,x,v)|^{p'}d\xi(x,v)dt\right)^{1\over p'}. \nonumber
\end{eqnarray}

In addition from \eqref{C11b}, H\"older inequality, \eqref{C12} and Lemma 4.2 it follows that
\begin{eqnarray}
&&\left|\int_0^T\int_{\Gamma_+}\psi(t,x,v){R_3^2}_{|(0,T)\times \Gamma_+}(t,x,v)d\xi(x,v)dt\right|\nonumber\\
&&=\left|\int_0^T\int_{\Gamma_+}\psi(t,x,v)\int_0^t\int_0^{t-t'}\int_{X\times\S^{d-1}}\tilde \gamma(t-t'-s_2,x,v,x',v')\right.\nonumber\\
&&\times\left.(U(s_2)A_2G_-(t')\phi)(x',v')dx'dv'ds_2dt'd\xi(x,v)dt\right|
\nonumber\\
&&=\left|\int_0^T\int_0^{T-t'}\int_{X\times\S^{d-1}}(U(s_2)A_2G_-(t')\phi)(x',v')\int_0^T\int_{\Gamma_+}\chi_{(t'+s_2,T)}(t)\psi(t,x,v)\right.\nonumber\\
&&\times\left.\tilde \gamma(t-t'-s_2,x,v,x',v')dtd\xi(x,v)dx'dv'ds_2dt'\right|\nonumber\\
&&\le \int_0^T\int_0^{T-t'}\int_{X\times\S^{d-1}}\left|U(s_2)A_2G_-(t')\phi\right|(x',v')\nonumber\\
&&\times\left(\int_0^T\int_{\Gamma_+}\chi_{(t'+s_2,T)}(t)|\tilde \gamma|^p(t-t'-s_2,x,v,x',v')dtd\xi(x,v)\right)^{1\over p}ds_2dt'dx'dv'\nonumber\\
&&\times\left(\int_{(0,T)\times\Gamma_+}|\psi(t,x,v)|^{p'}dtd\xi(x,v)\right)^{1\over p'}\nonumber\\
&&\le 2^{d-2}\|k\|_{\infty}^2\|\beta\|_{\infty}^{1\over p}\int_0^T\int_0^{T-t'}\int_{X\times\S^{d-1}}\left|U(s_2)A_2G_-(t')\phi\right|(x',v')dx'dv'ds_2dt\nonumber\\
&&\left(\int_{(0,T)\times\Gamma_+}|\psi(t,x,v)|^{p'}dtd\xi(x,v)\right)^{1\over p'}.\label{C18}
\end{eqnarray}
Moreover using \eqref{C15a}, the equality
$\|A_2\|=\|\sigma_p\|_\infty$ and the estimate $\|U(s_2)\|\le
e^{s_2\|\sigma_p\|_\infty}$ for $s_2\ge 0$ (see Trotter's formula [T]
$U(s_2)=s-\lim_{n\to \infty}\left(U_1({s_2\over n})e^{{s_2\over
      n}A_2}\right)^n$ where $U_1$ and $A_2$ are defined by \eqref{B4}
and \eqref{B2c} respectively), we obtain
\begin{eqnarray}
&&\int_0^T\int_0^{T-t'}\int_{X\times\S^{d-1}}\left|U(s_2)A_2G_-(t')\phi\right|(x',v')dx'dv'ds_2dt\nonumber\\
&\le&\int_0^T\int_0^{T-t'}{d\over ds_2}e^{s_2\|\sigma_p\|_\infty}ds_2dt' 
\|\phi\|_{L^1((0,\eta)\times \Gamma_-,dtd\xi)}\nonumber\\
&\le&T(e^{T\|\sigma_p\|_{\infty}}-1)\|\phi\|_{L^1((0,\eta),L^1( \Gamma_-,d\xi))}.\label{C18b}
\end{eqnarray}

Combining \eqref{C16b}--\eqref{C18b}, we finally obtain
\begin{eqnarray}
&&\left|\int_0^T\int_{\Gamma_+}\psi(t,x,v){(R_3^1+R_3^2)}_{|(0,T)\times \Gamma_+}(t,x,v)d\xi(x,v)dt\right|\nonumber\\
&&\le  2^{d-2}\|k\|_{\infty}^2\|\beta\|_{L^\infty(X)}^{1\over p}Te^{T\|\sigma_p\|_\infty}\|\phi\|_{L^1((0,\eta),L^1( \Gamma_-,d\xi))}\nonumber\\
&&\times\left(\int_0^T\int_{\Gamma_+}|\psi(t,x,v)|^{p'}d\xi(x,v)dt\right)^{1\over p'}.\label{C19}
\end{eqnarray}

Proposition 3.2 follows from \eqref{C4}, \eqref{C5} and \eqref{C19}.
\end{proof}

\section{Proof of Lemma 3.1, Theorems 3.1, 3.2} 
\begin{proof}[Proof of Lemma 3.1]
First note that using twice Lemma 4.1 we obtain
\begin{equation}
\int_{\Gamma_+}\int_0^{\tau_-(x,v)}\hspace{-.8cm}f(x-wv,v)dwd\xi(x,v)=\int_{X\times \S^{d-1}}\hspace{-.8cm}f(x,v)dxdv=\int_{\Gamma_-}\int_0^{\tau_+(x',v')}\hspace{-.8cm}f(x'+sv',v')dsd\xi(x',v'),\label{P1.0}
\end{equation}
for $f\in L^1(X\times \S^{d-1})$.\\

We first prove \eqref{P0h}. Let $T>\diam$. We have, in particular, $T>\tau_-(x,v)$ for any $(x,v)\in \bar X\times \S^{d-1}$.
From \eqref{P0f} and \eqref{P0c} it follows that
\begin{equation}
I_1(\psi,\ep_1,\ep_2)
=\int_{\Gamma_+}\Phi_{\ep_2}(x,v)f_{\ep_1}(x-\tau_-(x,v)v,v)d\xi(x,v),\label{P1.1}
\end{equation}
where $\Phi_{\ep_2}$ is the continuous function on $\Gamma_+$ given by
\begin{equation}
\Phi_{\ep_2}(x,v)=\int_{\tau_-(x,v)}^T\psi(t,x,v)g_{\ep_2}(t-\tau_-(x,v))dt\left(e^{-\int_0^{\tau_-(x,v)}\sigma(x-sv,v)ds}-e^{-\int_0^{\tau_-(x,v)}\tilde\sigma(x-sv,v)ds}\right),\label{P1.2}
\end{equation}
for $(x,v)\in \Gamma_+$ (we used also ${\rm
  supp}g_{\ep_2}\subseteq(0,+\infty)$). The continuity of
$\Phi_{\ep_2}$ follows from the assumptions \eqref{eq:hyp2p}, the
continuity of $\psi$ and $g_{\ep_2}$ and the continuity of $\tau_-$ on
$\Gamma_+$ ($X$ is convex with $C^1$ boundary).  From \eqref{P1.0}
(``$f(x,v)={1\over
  \tau(x,v)}\Phi_{\ep_2}(x+\tau_+(x,v)v,v)f_{\ep_2}(x-\tau_-(x,v)v,v)$'')
and \eqref{P0a}, we obtain
\begin{eqnarray}
&&\int_{\Gamma_+}\!\!\!\!\!\Phi_{\ep_2}(x,v)f_{\ep_1}(x-\tau_-(x,v)v,v)d\xi(x,v)
=\int_{\Gamma_-}\!\!\!\!\!\Phi_{\ep_2}(x'+\tau_+(x',v')v',v')f_{\ep_1}(x',v')d\xi(x',v')\nonumber\\
&&\underset{\ep_1\to 0^+}{\longrightarrow}\int_{\tau_+(x_0',v_0')}^T\psi(t,x_0'+\tau_+(x_0',v_0')v_0',v_0')
\left(e^{-\int_0^{\tau_+(x_0',v_0')}\sigma(x_0'+sv_0')ds}-e^{-\int_0^{\tau_+(x_0',v_0')}\tilde\sigma(x_0'+sv_0')ds}\right)\nonumber\\&&\hspace{2cm}\times
g_{\ep_2}(t-\tau_+(x_0',v_0'))dt\label{P1.3}.
\end{eqnarray}
The limit \eqref{P0h} follows from \eqref{P1.1}, \eqref{P1.3}, \eqref{P0b} and the continuity of $\psi$ and \eqref{eq:hyp2p}.\\

We prove \eqref{P0i}. Let $T>2\diam$. We have, in particular,
$T>s+\tau_+(x'+sv',v)$ for $s\in (0,\tau_+(x',v'))$ and $(x',v')\in
\Gamma_-$. From \eqref{P0g} , it follows that
\begin{eqnarray}
&&I_2(\psi,\ep_1,\ep_2) =\int_{\Gamma_+}\int_0^{\tau_-(x,v)}\int_0^T\psi(t,x,v)\label{P1.4}\\&&\times\int_{\S^{d-1}}(k(x-wv,v',v)E_-(x,v,w,v')-\tilde k(x-wv,v',v)\tilde E_-(x,v,w,v'))\nonumber\\
&&\times g_{\ep_2}(t-w-\tau_-(x-wv,v'))f_{\ep_1}(x-wv-\tau_-(x-wv,v')v',v')dv'dtdwd\xi(x,v).\nonumber
\end{eqnarray}
Using \eqref{P1.4} and \eqref{P1.0}, we obtain
\begin{equation}
I_2(\psi,\ep_1,\ep_2)=\int_{\Gamma_-}\Psi_{\ep_2}(x',v') f_{\ep_1}(x',v')d\xi(x',v'),\label{P1.5}
\end{equation}
where
\begin{eqnarray}
&&\Psi_{\ep_2}(x',v')=\int_{\S^{d-1}}\int_0^{\tau_+(x',v')}\int_{s+\tau_+(x'+sv',v)}^T\psi(t, x'+sv'+\tau_+(x'+sv',v)v,v)\nonumber\\
&&\times\left(k(x'+sv',v',v)E_+(x',v',s,v)-\tilde k(x'+sv',v',v)\tilde E_+(x',v',s,v)\right)\nonumber\\
&&\times g_{\ep_2}(t-s-\tau_+(x'+sv',v))dt ds dv,\label{P1.6}
\end{eqnarray}
for $(x',v')\in \Gamma_-$. From \eqref{eq:hyp2p}, \eqref{P-1} and the
continuity of $\psi$ and $g_{\ep_2}$, it follows that $\Psi_{\ep_2}$
is continuous on $\Gamma_-$. From \eqref{P0a} and \eqref{P1.6} it
follows that
\begin{eqnarray}
&&\int_{\Gamma_-}\Psi_{\ep_2}(x',v') f_{\ep_1}(x',v')d\xi(x',v')\nonumber\\
&&\underset{\ep_1\to 0^+}{\longrightarrow}\int_{\S^{d-1}}\int_0^{\tau_+(x_0',v_0')}\left(\int_{s+\tau_+(x_0'+sv_0',v)}^T\psi(t, x_0'+sv_0'+\tau_+(x_0'+sv_0',v)v,v)\right.\nonumber\\
&&(E_+(x_0',v_0',s,v)k(x_0'+sv_0',v_0',v)-\tilde E_+(x_0',v_0',s,v)\tilde k(x_0'+sv_0',v_0',v))
\label{P1.7}\\
&&\times \left.g_{\ep_2}(t-s-\tau_+(x_0'+sv_0',v))dt\right) ds dv.\nonumber
\end{eqnarray}
The limit \eqref{P0i} follows from \eqref{P1.5}, \eqref{P1.7},
\eqref{P0b}, \eqref{eq:hyp2p}, the continuity of $\psi$ and the
Lebesgue dominated convergence theorem.
\end{proof}

\begin{proof}[Proof of Theorem 3.1]
We first prove \eqref{P0k}. Let $T>\diam$. Let $\ep_3>0$ and let $\psi_{\ep_3}$ be a continuous and compactly supported function on $(0,T)\times \Gamma_+$ that satisfies
\begin{eqnarray}
&&0\le \psi_{\ep_3}\le 1\textrm{ and } {\rm supp}\psi_{\ep_3}\subseteq\{(t,x,v)\in (0,T)\times \Gamma_+\ |\ |v-v_0'|<\ep_3\},\label{P2}\\
&&\psi_{\ep_3}(t,x,v)=1 \textrm{ for }(t,x,v)\in (0,T)\times \Gamma_+\textrm{ such that}\nonumber\\
&&|v-v_0'|\le{\ep_3\over 2}, |t-\tau_+(x_0',v_0')|\le{T-\tau_+(x_0',v_0')\over 2}. \nonumber
\end{eqnarray}

From \eqref{P0h} and \eqref{P2}, it follows that
\begin{equation}
I_1(\psi_{\ep_3},\ep_1,\ep_2)=e^{-\int_0^{\tau_+(x_0',v_0')}\sigma(x_0'-sv_0',v_0')ds}-e^{-\int_0^{\tau_+(x_0',v_0')}\tilde \sigma(x_0'-sv_0',v_0')ds}.\label{P3a}
\end{equation} 
 
Using \eqref{P0i}, \eqref{P2} and  the estimate $\sigma \ge 0$, we obtain 
\begin{equation*}
\left|\lim_{\ep_2\to 0^+}\lim_{\ep_1\to 0^+}I_2(\psi_{\ep_3},\ep_1,\ep_2)\right|\le{\rm diam}(X)\left(\|k\|_{\infty}+\|\tilde k\|_{\infty}\right)\int_{v\in\S^{d-1}\atop |v-v_0'|<\ep_3}dv.
\end{equation*}
Hence
\begin{equation}
\lim_{\ep_3\to 0^+}\lim_{\ep_2\to 0^+}\lim_{\ep_1\to 0^+}I_2(\psi_{\ep_3},\ep_1,\ep_2)=0.\label{P3b}
\end{equation}

From \eqref{P0gg} and \eqref{P2}, it follows that
\begin{eqnarray*}
\left|I_3(\psi_{\ep_3},\ep_1,\ep_2)\right|
&\le& C\Big(\int_0^T\int_{\pa X}\int_{v\in \S^{d-1}\atop |v-v_0'|<\ep_3}dv d\mu(x)dt\Big)^{1\over p'}\\
&\le&C ({\rm Vol}(\pa X)T)^{1\over p'}\Big(\int_{v\in\S^{d-1}\atop |v-v_0'|<\ep_3}dv\Big)^{1\over p'},
\end{eqnarray*}
for $\ep_i>0$, $i=1\ldots 3$. Therefore
\begin{equation}
\lim_{\ep_3\to 0^+}\limsup_{\ep_2\to 0^+}\limsup_{\ep_1\to 0^+}I_3(\psi_{\ep_3},\ep_1,\ep_2)=0.\label{P3c}
\end{equation}

In addition, from \eqref{P0d}--\eqref{P0e} it follows that
\begin{equation}
|I_1(\psi_{\ep_3},\ep_1,\ep_2)|\le \|A-\tilde A\|_{\eta,T}+|I_2(\psi_{\ep_3},\ep_1,\ep_2)+I_3(\psi_{\ep_3},\ep_1,\ep_2)|,\label{P3d}
\end{equation}
for $\ep_i>0$, $i=1\ldots 3$.

Combining \eqref{P3a}--\eqref{P3d} we obtain \eqref{P0k}.\\

Now we prove \eqref{P0l}. Let $T>2\diam$.  Let $U:=\{(t',v)\in
(0,\tau_+(x_0',v_0'))\times\S^{d-1}\ |\ (k-\tilde
k)(x_0'+t'v_0',v_0',v)>0\}$. From \eqref{eq:hyp2p} it follows that $U$
is an open subset of $\R\times \S^{d-1}$.  Let $(K_m)$ be a sequence
of compact sets such that $\bigcup_{m\in \N}K_m=U$ and $K_m\subseteq
K_{m+1}$ for $m\in \N$.  For $m\in \N$ let $\chi_m\in
C^\infty(\R\times \S^{d-1},\R)$ such that $\chi_{K_m}\le \chi_m\le
\chi_U$ (where $\chi_{K_m}$ and $\chi_U$ are defined in \eqref{D4-5}),
and let
\begin{equation}
\rho_m=2\chi_m-1.\label{P4a}
\end{equation}
Thus we obtain
\begin{equation}
\lim_{m\to +\infty}(k-\tilde k)(x_0'+t'v_0',v_0',v)\rho_m(t',v)=
|k-\tilde k|(x_0'+t'v_0',v_0',v),\label{P4b}
\end{equation}
for $v\in \S^{d-1}$ and $t'\in (0,\tau_+(x_0',v_0'))$.

Consider
\begin{equation}
{\mathcal V}_\delta:=  \{(t,x,v)\in (0,T)\times\Gamma_+\ | \ |v-(vv_0')v_0'|>\delta,\ {\delta\over 2}<t<T-{\delta\over 2}\},\quad\label{P4e}
\end{equation}
\begin{equation}
{\mathcal V}_{\delta,l}:=\{(t,x,v)\in (0,T)\times\Gamma_+\ |\ | v-(vv_0')v_0'|\ge\delta+{1\over l},\ \delta\le t\le T-\delta\},\label{P4f}
\end{equation}
for $0<\delta<{\min(1,T)\over 2}$ and $l\in \N$, $l\ge 2$.  For $0<\delta<{\min(1,T)\over 2}$ and $l\in \N$, $l\ge 2$, let $\chi_{\delta,l}$ be a continuous and compactly 
supported function on $(0,T)\times\Gamma_+$ such that
\begin{equation}
\chi_{{\mathcal V}_{\delta,l}}\le \chi_{\delta,l}\le \chi_{{\mathcal V}_\delta}\label{P4g}
\end{equation}
(where $\chi_{{\mathcal V}_{\delta,l}}$ and $\chi_{{\mathcal
    V}_\delta}$ are defined in \eqref{D4-5}).  Finally, for
$0<\delta<{\min(1,T)\over 2}$ and $m,l\in \N$, $l\ge 2$, let
$\psi_{\delta,m,l,\ep_3}$ be the continuous compactly supported
function on $(0,T)\times \Gamma_+$ defined by
\begin{equation}
\psi_{\delta,m,l,\ep_3}(t,x,v):=\chi_{\delta,l}(t,x,v)\left(\zeta_{\ep_3}(t-s-s')\rho_m(s,v)\right)_{s={(x-x_0')(v_0'-( vv_0') v)\over 1-( vv_0')^2}
\atop s'={(x-x_0')( v-(vv_0')v_0')\over 1-(vv_0')^2}},\label{P4h}
\end{equation}
where $\zeta_{\ep_3}\in C^\infty(\R)$, $\zeta_{\ep_3}(s'')=1$ for $s''\in [-\ep_3,\ep_3]$, $0\le \zeta_{\ep_3}\le 1$ and $\zeta_{\ep_3}(s'')=0$ for $|s''|\ge 2\ep_3$. 

From \eqref{P0h}, \eqref{P4h} and the equality $\chi_{\delta,l}(t,x_0'+\tau_+(x_0',v_0')v_0', v_0')=0$ for $t\in (0,T)$ (see \eqref{P4e}--\eqref{P4g}), it follows that
\begin{equation}
\lim_{\ep_2\to0^+}\lim_{\ep_1\to 0^+}I_1(\psi_{\delta,m,l,\ep_3}, \ep_1,\ep_2)=0\label{P5a}
\end{equation}
for $0<\delta<{\min(1,T)\over 2}$, $m$, $l\ge 2$, $\ep_3>0$.

From \eqref{P0ja}--\eqref{P0jb} and  \eqref{P4h}, it follows that 
\begin{eqnarray}
I_2^1(\psi_{\delta,m,l,\ep_3})&:=&\int_{\S^{d-1}}\int_0^{\tau_+(x_0',v_0')}\zeta_{\ep_3}(0)\chi_{\delta,l}(s+\tau_+(x_0'+sv_0',v),x_0'+sv_0'+\tau_+(x_0'+sv_0',v)v,v)\nonumber\\
&&\times \rho_m(s,v)
(k-\tilde k)(x_0'+sv_0',v_0',v)E_+(x_0',v_0',s,v) ds dv.\label{P5c}\\
I_2^2(\psi_{\delta,m,l,\ep_3})&:=&\int_{\S^{d-1}}\int_0^{\tau_+(x_0',v_0')}\zeta_{\ep_3}(0)\chi_{\delta,l}(s+\tau_+(x_0'+sv_0',v),x_0'+sv_0'+\tau_+(x_0'+sv_0',v)v,v)\nonumber\\
&&\times \rho_m(s,v)\tilde k(x_0'+sv_0',v_0',v)\left(E_+-\tilde E_+\right)(x_0',v_0',s,v)ds dv,\label{P5d}
\end{eqnarray}
for $0<\delta<{\min(1,T)\over 2}$, $m$, $l\ge 2$, $\ep_3>0$.

Note that using \eqref{P4e}--\eqref{P4g} we obtain 
\begin{equation}
\lim_{l\to\infty}\chi_{{\mathcal V}_{\delta,l}}(t,x,v)=\chi_{{\mathcal V}_\delta}(t,x,v), \label{P5f}
\end{equation}
for $(t,x,v)\in (0,T)\times\Gamma_+$ and $0<\delta<{\min(1,T)\over 2}$.

From equality $\zeta_{\ep_3}(0)=1$, \eqref{P5c}, \eqref{P5f} and the
Lebesgue dominated convergence theorem, it follows that
\begin{eqnarray*}
\lim_{l\to +\infty}I_2^1(\psi_{\delta,m,l,\ep_3})&=&\int_{\S^{d-1}}\int_0^{\tau_+(x_0',v_0')}\chi_{{\mathcal V}_\delta}(s+\tau_+(x_0'+sv_0',v),x_0'+sv_0'+\tau_+(x_0'+sv_0',v)v,v)\nonumber\\
&&\times \rho_m(s,v)(k-\tilde k)(x_0'+sv_0',v_0',v)E_+(x_0',v_0',s,v) ds dv,
\end{eqnarray*}
for $0<\delta<{\min(1,T)\over 2}$, $m\in \N$, $l\ge 2$, $\ep_3>0$.
Therefore, using \eqref{P4b} and the Lebesgue dominated convergence
theorem, we obtain
\begin{eqnarray*}
\lim_{m\to +\infty}\lim_{l\to +\infty}I_2^1(\psi_{\delta,m,l,\ep_3})
&=&\int_{\S^{d-1}}\int_0^{\tau_+(x_0',v_0')}\hspace{-1cm}\chi_{{\mathcal V}_\delta}(s+\tau_+(x_0'+sv_0',v),x_0'+sv_0'+\tau_+(x_0'+sv_0',v)v,v)\nonumber\\
&&\times E_+(x_0',v_0',s,v)|k-\tilde k|(x_0'+sv_0',v_0',v)ds dv,
\end{eqnarray*} 
for $0<\delta<{\min(1,T)\over 2}$, $\ep_3>0$.
Using this latter equality and \eqref{P4e}, we obtain
\begin{equation}
\lim_{\delta\to 0^+}\lim_{\ep_3\to 0^+}\lim_{m\to +\infty}\lim_{l\to +\infty}I_2^1(\psi_{\delta,m,l,\ep_3})
=\int_{\S^{d-1}}\int_0^{\tau_+(x_0',v_0')}\hspace{-.25cm}|k-\tilde k|(x_0'+sv_0',v_0',v) E_+(x_0',v_0',s,v)ds dv.\label{P5h}
\end{equation}

From equality $\zeta_{\ep_3}(0)=1$ and \eqref{P5d} it follows that
\begin{equation}
|I_2^2(\psi_{\delta,m,l,\ep_3})|\le{\rm diam}(X)\sup_{s\in (0,\tau_+(x_0',v_0'))}\tilde \sigma_p(x_0'+sv_0',v_0')
\sup_{v\in \S^{d-1}\atop 0<s<\tau_+(x_0',v_0')}\left|(E_+-\tilde E_+)(x_0',v_0',s,v)\right|,\label{P5i}
\end{equation}
for $0<\delta<{\min(1,T)\over 2}$, $m\in \N$, $l\ge 2$, $\ep_3>0$.

Note that using \eqref{P4h} and the estimate $0\le \rho_m\le 1$ for
all $m$, we obtain
\begin{equation}
\begin{array}{ll}
&\displaystyle\int_0^T\int_{\Gamma_+}|\psi_{\delta,m,l,\ep_3}(t,x,v)|^{p'}d\xi(x,v)dt\\
\le& \displaystyle\int_0^T\int_{\Gamma_+}\chi_{\delta,l}(t,x,v)^{p'}
\zeta_{\ep_3}(t-s-s')^{p'}_{s={(x-x_0')(v_0'-( vv_0') v)\over 1-( vv_0')^2}\atop s'={(x-x_0')( v-(vv_0')v_0')\over 1-(vv_0')^2}}
\end{array}
d\xi(x,v)dt,\label{P5j}
\end{equation}
for $0<\delta<{\min(1,T)\over 2}$, $m\in \N$, $l\ge 2$, $\ep_3>0$.
Therefore using the definition of $\zeta_{\ep_3}$ we obtain
\begin{equation}
\int_0^T\int_{\Gamma_+}|\psi_{\delta,m,l,\ep_3}(t,x,v)|^{p'}d\xi(x,v)dt\le \int_0^T\!\!\!\int\limits_{\Gamma_+\atop |v-(vv_0')v_0'|>\delta}\!\!\!\!\!\!\!\!\!\!\!
\!
\chi_{[-2\ep_3,2\ep_3]}(t-s-s')_{s={(x-x_0')(v_0'-( vv_0') v)\over 1-( vv_0')^2}\atop s'={(x-x_0')( v-(vv_0')v_0')\over 1-(vv_0')^2} }
d\xi(x,v)dt,\label{P5k}
\end{equation}
for $0<\delta<{\min(1,T)\over 2}$, $m\in \N$, $l\ge 2$, $\ep_3>0$.
Using \eqref{P5j}--\eqref{P5k} and the Lebesgue dominated convergence
theorem, we obtain
\begin{equation*}
\lim_{\ep_3\to 0^+}\limsup_{m\to +\infty}\limsup_{l\to +\infty}\int_0^T\int_{\Gamma_+}|\psi_{\delta,m,l,\ep_3}(t,x,v)|^{p'}d\xi(x,v)dt=0,
\end{equation*}
for $0<\delta<{\min(1,T)\over 2}$.
Using this latter equality and \eqref{P0gg}, we obtain
\begin{equation}
\lim_{\delta\to 0^+}\lim_{\ep_3\to 0^+}\limsup_{m\to+\infty}\limsup_{l\to +\infty}\lim\sup_{\ep_2\to 0^+}\limsup_{\ep_1\to 0^+}
\left|I_3(\psi_{\delta,m,l,\ep_3},\ep_1,\ep_2)\right|
=0.\label{P5l}
\end{equation}
In addition, from \eqref{P0d}--\eqref{P0e},  it follows that
\begin{equation}
|I_2(\psi_{\delta,m,l,\ep_3},\ep_1,\ep_2)|\le \|A-\tilde A\|_{\eta,T}+|I_1(\psi_{\delta,m,l,\ep_3},\ep_1,\ep_2)+I_3(\psi_{\delta,m,l,\ep_3},\ep_1,\ep_2)|,\label{P5m}
\end{equation}
for $0<\delta<{\min(1,T)\over 2}$, $m\in \N$, $l\ge 2$, $\ep_i>0$, $i=1\ldots 3$.
Combining \eqref{P5m}, \eqref{P5h}, \eqref{P5i} and \eqref{P5l} we obtain \eqref{P0l}.
\end{proof}

\begin{proof}[Proof of Theorem 3.2]
  The method used to prove \eqref{t5.2a} is the same as in [W] and
  [BJ1].  For the reader's convenience, we adapt the proof given in
  [BJ1] with minor modification.

Let $(\sigma,k)$, $(\tilde \sigma,\tilde k)\in
{\mathcal M}$. We extend $\sigma$ and $\tilde \sigma$ outside $X$ by $0$. Let $f=\sigma-\tilde\sigma$ and consider $Pf$ the X-ray
transform of $f=\sigma-\tilde\sigma$ defined by
$Pf(x,\varphi):=\int_{-\infty}^{+\infty}f(t\varphi+x)dt$ for 
$(x,\varphi)\in T\S^{d-1}:=\{(z,v)\in \R^d\times \S^{d-1}\ |\ vz=0\}$.

From $f_{|X}\in H^{{d\over 2}+\tilde r}(X)$, it follows that
\begin{equation}
\|f\|_{H^{-{1\over 2}}(X)}\le D_1(d,X)\|Pf\|_{*},\label{t3.2p1}
\end{equation}
where 
$$
\|Pf\|_{*}:=\left(\int_{\S^{d-1}}\int_{\Pi_\varphi}|Pf(x,\varphi)|^2dx d\varphi\right)^{1\over 2}
$$
and $D_1(d,X)$ is a real constant which does not depend on $f$ and  $\Pi_\varphi:=\{x\in \R^d\ |\ x\varphi=0\}$  for $\varphi\in \S^{d-1}$.
Note that $Pf(x,\varphi)=0$ for $(x,\varphi)\in T\S^{d-1}$ and $|x|\ge \sup_{z\in X}|z|$. Therefore using also \eqref{t3.2p1} we obtain 
\begin{equation}
\|f\|_{H^{-{1\over 2}}(X)}\le D_2(d,X)\|Pf\|_{L^\infty(T\S^{d-1})},\label{t3.2p2}
\end{equation}
where $D_2(d,X)$ is a real constant which does not depend on $\sigma$, $\tilde \sigma$.

We also use the following interpolation inequality:
\begin{equation}
\|f\|_{H^s(X)}\le \|f\|_{H^{{d\over 2}+\tilde r}}^{2s+1\over d+1+2\tilde r}\|f\|_{H^{-{1\over 2}}}^{d+2\tilde r\over d+1+2\tilde r},\label{t3.2p2b}
\end{equation}
for $-{1\over 2}\le s\le {d\over 2}+\tilde r$.  As $(\sigma,k)\in {\mathcal M}$, it follows that
\begin{equation}
\|\sigma\|_\infty\le D_3(d,\tilde r)\|\sigma\|_{H^{{d\over 2}+\tilde r}}\le D_3(d,\tilde r)M.\label{t3.2p3}
\end{equation}
Therefore,
\begin{equation}
\int_0^{\tau_+(x_0',v_0')}\sigma (x_0'+sv_0')ds\le \diam D_3(d,\tilde r)M,\label{t3.2p4}
\end{equation}
for  $(x_0',v_0')\in \Gamma_-$.  From \eqref{t3.2p4}, it
follows that
\begin{equation}
\left|e^{-\int_0^{\tau_+(x_0',v_0')}\sigma (x_0'+sv_0',v_0')ds }
-e^{-\int_0^{\tau_+(x_0',v_0')}\tilde\sigma (x_0'+sv_0',v_0')ds }\right|\ge e^{-\diam D_3(d,\tilde r)M}|P(\sigma-\tilde \sigma)(x_0',v_0')|,\label{t3.2p5}
\end{equation}
for $(x_0', v_0')\in \Gamma_-$ (we used the equality $e^{t_1}-e^{t_2}=e^{c}(t_2-t_1)$ for $t_1<t_2\in \R$ and for some $c\in [t_1,t_2]$, which depends 
on $t_1$ and $t_2$).

Combining \eqref{t3.2p5}, \eqref{t3.2p2} and
\eqref{P0k}, we obtain
\begin{equation}
{e^{- \diam D_3(d,\tilde r)M}\over D_2(d,X)}\|\sigma-\tilde \sigma\|_{H^{-{1\over 2}}(X)}\le \|A-\tilde A\|_{\eta,T}.\label{t3.2p6}
\end{equation}
Combining \eqref{t3.2p6} and \eqref{t3.2p2b}, we obtain \eqref{t5.2a}.

We now prove \eqref{t5.2b}.  Using \eqref{P-1} and \eqref{t3.2p3}, we obtain that
$$
\int_{\S^{d-1}}\int_0^{\tau_+(x_0',v_0')}\left|k(x_0'+sv_0',v_0',v)-\tilde
  k(x_0'+sv_0',v_0',v)\right|
E(x_0',v_0',s,v)dsdv
$$
\begin{equation}
\ge e^{-2\diam D_3(d,\tilde r)M}\int_{\S^{d-1}}\int_0^{\tau_+(x_0',v_0')}|k-\tilde k|(x_0'+sv_0',v_0',v)|dsdv,\label{t3.2p7}
\end{equation}
for any $(x_0',v_0')\in \Gamma_-$.

As $(\tilde \sigma,\tilde k)\in {\mathcal M}$ we have $\|\tilde
\sigma_p\|_{\infty}\le M$. Using the latter estimate and \eqref{P-1}, we obtain
\begin{eqnarray}
&&\sup_{s\in (0,\tau_+(x_0',v_0'))}\sigma_p(x_0'+sv_0',v_0')\sup\limits_{(x_0',v_0')\in \Gamma_-\atop s\in (0,\tau_+(x_0',v_0'))}|E-\tilde E|(x_0',v_0',s,v)\le Me^{2\diam D_3(d,\tilde r)M}\nonumber\\
&&\times\sup\limits_{(x_0',v_0')\in \Gamma_-\atop s\in (0,\tau_+(x_0',v_0'))}\left[\int_0^s\!\!\!\!\!|\sigma-\tilde \sigma|(x-pv,v)dp+
\int_0^{\tau_+(x_0'+sv_0',v)}\!\!\!\!\!\!\!\!\!\!\!\!\!\!\!\!\!\!\!|\sigma-\tilde \sigma|(x_0'+sv_0'+pv,v)dp\right]\nonumber
\end{eqnarray}
\begin{equation}
\le 2\diam Me^{2\diam D_3(d,\tilde r)M}\|\sigma-\tilde \sigma\|_{\infty},\label{t3.2p8}
\end{equation}
for any $(x_0',v_0')\in \Gamma_-$. (We also used $|e^u-e^{\tilde
  u}|\le e^{\max(|u|, |\tilde u|)}|u-\tilde u|$ where
$u=-\int_0^s\sigma(x_0'+pv_0',v_0')
dp-\int_0^{\tau_+(x_0'+sv_0',v)}\tilde\sigma(x_0'+sv_0'+pv,v)dp$ and $\tilde u$ denotes the real number
obtained by replacing $\sigma$ by $\tilde \sigma$ on the right-hand
side of the latter equality that defines $u$ ; using \eqref{t3.2p3}
(for $\sigma$ and for $\tilde \sigma$) we obtain $\max(|u|,|\tilde
u|)\le 2\diam D_3(d,\tilde r)M$.)  Note that
$\|\sigma-\tilde\sigma\|_\infty\le D_3(d,r)\|\sigma-\tilde
\sigma\|_{H^{{d\over 2}+r}}$ for $0<r<\tilde r$ (see \eqref{t3.2p3}).
Therefore, combining \eqref{t3.2p7}, \eqref{t3.2p8}, \eqref{P0l} and
\eqref{t5.2a}, we obtain \eqref{t5.2b}.

Let us finally prove \eqref{t5.2c}. Let $0<r<\tilde r$ and let
$\kappa={2(\tilde r-r)\over d+1+2 \tilde r}$. From \eqref{t5.2b} it
follows that
\begin{equation}
\int_{\Gamma_-}\int_0^{\tau_+(x_0',v_0')}\int_{\S^{d-1}}\left|(k-\tilde k)(x_0'+sv_0',v_0',v)\right|dvdsd\xi(x_0',v_0')
\le D_4\|A-\tilde A\|_{\eta,T}^\kappa
\left(1+\|A-\tilde A\|_{\eta,T}^{1-\kappa}\right),\label{t3.2p9}
\end{equation}
where $D_4=C_2\int_{\Gamma_-}d\xi(x_0',v_0')$ and $C_2$ is the
constant that appears on the right-hand side of \eqref{t5.2b}.  From
\eqref{t3.2p9} and Lemma 4.1, we obtain \eqref{t5.2c}.
\end{proof}

\section*{Acknowledgments}
This work was funded in part by the National Science Foundation under
Grants DMS-0554097 and DMS-0804696.

\section*{References}
\begin{itemize}
\item[{[BJ1]}]G. Bal and A. Jollivet, {\it Stability estimates in
    stationary inverse transport}, 2008 preprint, arXiv:0804.1320.

\item[{[BJ2]}]G. Bal and A. Jollivet, {\it in preparation}.
  
\item[{[BLM]}]G. Bal, I. Langmore, and F. Monard, {\em Inverse
    transport with isotropic sources and angularly averaged
    measurements}.  \newblock{Inverse Probl. Imaging}, {\bf 2}:1,
  23--42 (2008).
  
\item[{[C1]}]M. Cessenat, {\it Th\'eor\`emes de trace $L^p$ pour des
    espaces de fonctions de la neutronique}, C. R. Acad. Sci. Paris
  Sér. I Math.  {\bf 299}:16, 831--834 (1984).
  
\item[{[C2]}]M. Cessenat, {\it Th\'eor\`emes de trace pour des espaces
    de fonctions de la neutronique}, C. R. Acad. Sci. Paris Sér. I
  Math.  {\bf 300}:3, 89--92 (1985).
  
\item[{[CMR]}]R. Cipolatti, C.M. Motta and N.C. Roberty, {\it
    Stability estimates for an inverse problem for the linear
    Boltzmann equation}, Rev. Mat. Complut. {\bf 19}:1, 113--132
  (2006).
  
\item[{[CS1]}]M. Choulli and P. Stefanov, {\it Inverse scattering and
    inverse boundary value problems for the linear Boltzmann
    equation}, Comm. P.D.E. {\bf 21}, 763--785 (1996).
  
\item[{[CS2]}]M. Choulli and P. Stefanov, {\it An inverse boundary
    value problem for the stationary transport equation}, Osaka J.
  Math. {\bf 36}, 87--104 (1999).
  
\item[{[DL]}]R. Dautray and J.-L. Lions, {\it Mathematical Analysis
    and Numerical Methods for Science and Technology}, Vol. 6, Sringer
  Verlag, Berlin, 1993.
  
\item[{[L]}]I. Langmore, {\em The stationary transport equation with
    angularly averaged measurements}, Inverse Problems, {\bf 24},
  015924 (2008).
  
  
\item[{[R1]}]V.~G. Romanov, {\em Estimation of stability in the
    problem of determining the attenuation coefficient and the
    scattering indicatrix for the transport equation}, (Russian)
  Sibirsk. Mat. Zh., {\bf 37}:2, 361--377 (1996); translation in
  Siberian Math. J. {\bf 37}:2,  308--324 (1996).

\item[{[R2]}]V.~G. Romanov, 
  {\em Stability estimates in the three-dimensional inverse
  problem for the transport equation}, J. Inverse Ill-Posed Probl. {\bf 5},
  463--475 (1997).

\item[{[S]}]P.~Stefanov, {\em Inside Out: Inverse problems and
    applications}, vol.~47 of MSRI publications, Ed. G. Uhlmann,
  Cambridge University Press, Cambridge, UK, 2003, ch.~Inverse
  Problems in Transport Theory.
  
\item[{[SU]}] P.~Stefanov and G.~Uhlmann, {\em Optical tomography in
    two dimensions}, Methods Appl. Anal., {\bf 10}, 1--9 (2003).
  
\item[{[T]}]H.F. Trotter, {\it On the product of semi-groups of
    operators}, Proc. Am. Math. Soc. {\bf 10}:4, 545--551 (1959).
  
\item[{[W]}]J. Wang, {\it Stability estimates of an inverse problem
    for the stationary transport equation}, Ann. Inst. H. Poincar\'e
  Phys. Th\'eor. {\bf 70}:5, 473--495 (1999).
\end{itemize}
\end{document}